\documentstyle{article}
\begin{document}

\noindent {\bf Bateman method for two-body scattering without partial-wave decomposition}

  \begin{center}
  Zeki C. Kuruo\u{g}lu\\
  {\it Department of Chemistry, Bilkent University,
 06800 Bilkent, Ankara,Turkey}
  \end{center}

  We explore the use of Bateman method for solving
  the two-variable version of the Lippmann-Schwinger
  equation for two-body scattering without invoking  the partial-wave decomposition.
   In our adaptation of the Bateman method to the Lippmann-Schwinger equation,
  the momentum-space kernel  of
   the potential  is interpolated by a separable expansion constructed
   from its sections on a multi-variate grid.  We describe a suitable scheme
   for constructing  a  multi-variate Cartesian grid  that allows  for
   the  treatment of the singularity of the free propagator with due care.
  The method is tested  in  the nucleon-nucleon scattering
   employing   a model two-nucleon potential.
   Our results demonstrate   that the Bateman method can produce
   quite accurate solutions with relatively small number of grid points.\\


\vspace*{1cm}

\begin {center} {\bf I. INTRODUCTION } \end{center}

The traditional strategy in the treatments of quantum  scattering problems  have been the
 elimination of angular variables via expansions over angular-momentum states.
 Certain drawbacks of this strategy have been noted in recent years, especially for high-energy
collisions and within  the context of few-body problems. As a result, computational
methods that avoid the traditional decomposition of wave functions
and scattering amplitudes into partial waves have been explored recently
by a number of groups [1-13].
 Various direct multi-variable solution techniques for
 two-body Lippmann Schwinger (LS) equation have been investigated.
 Most studies employed   the Nystrom method (i.e.,
  discretization of the integral equation via a suitable multi-variate quadrature)[14].
  Although the Nystrom method   can produce very accurate
   results, the matrix dimensions in the multi-variable Nystrom approach
    can grow very fast to computationally prohibitive levels.
     Multi-variable methods
   that lead to a reduction in the matrix sizes
   are therefore of considerable interest.

   Methods based on basis-set expansions, such as  Galerkin [3,5], collocation [13], and
   Schwinger variational methods [13],  have been
   investigated with various choices of multi-variable bases.
  These  expansion methods give rise to  matrix elements involving multi-variable
  integrals that are typically evaluated by quadrature.
  Assuming that the  quadrature rule used in Nystrom method
   is also employed to construct the matrix elements of,
    say, the Galerkin method,
   the final linear equation system of the Galerkin method
   can  be viewed as a contraction of the linear system of equations
   that  the Nystrom approach gives rise to.
    In effect, the basis-set expansion methods represent
  replacement of the (large) equation system of the Nystrom method
  by a smaller set of approximate equations,
   by demanding that a residual vanishes on a chosen test space [13].
   In numerical linear algebra,  such contractions/projections
   form the basis for  many (iterative) methods to solve
   large systems of equations.

   In this article we consider the Bateman method
   as a tool for the solution of the multi-variable integral equations
   of the two-body scattering problem.
   This method was originally proposed by Bateman [15]
   to solve single-variable integral equations by
   interpolating the bi-variate kernel by its sections  on an interpolation  grid.
   Analysis of its convergence and error bound is given in Refs. [16] and [17].

   The Bateman method  had drawn some interest in 1970's
   in the context of few-body
   scattering calculations[18-23]. However, almost all discussions and
   applications   have been restricted to partial-wave (single-variable)
   LS equations. As far as the present author is aware,
   the only Bateman application
    that avoids partial-wave expansion is by
    Gianini and Lim [22] who applied  the Bateman method  to  the  two-variable
   LS equation for a Yukawa potential.
   The interpolation grid in their implementation
   had been severely limited, and
   as a result the full potential of the method
   does not seem to have been realized.

     Although the Bateman method could be used directly
     on the three-dimensional LS equation,
     the present paper concentrates on its application
     to  the reduced (two-variable) version of the LS equation
     in which the azimuthal angle dependence is integrated out.
     We use a  cartesian bi-variate grid
     that  pays due care  for a symmetric treatment of the
    singularity of the free propagator in the kernel of the LS equation.
    The results of our calculations
     for the Malfliet-Tjon potential [24] demonstrate that
    the promise of the Bateman method
     as a simple and viable tool to solve multi-variable
     scattering equations is in fact borne out.
     It appears that the choice for the
     interpolation grid is instrumental in obtaining
     rather accurate results
     with relatively few interpolation points.

      Another observation that does not appear to have been noticed before  is
     the intimate relation between Nystrom and Bateman
     methods:  A  Nystrom  calculation (with a given quadrature grid)
     can also be viewed as a Bateman calculation in which
     the quadrature grid employed in  the Nystrom method serves   as both
     interpolation and quadrature grids.
     This equivalence is borne out in our numerical calculations as well,
     even though  our implementations
      of Nystrom and Bateman methods are organized differently.
      When the quadrature grid used in Nystrom calculation is employed as both interoplation and
      quadrature grid in the Bateman calculation, the results of two calculations are indistinguishable
      within at least 8 digits after the decimal point.

          Plan of this article is as follows: In Sect. II, we fix notation and give a derivation
            of the reduced  two-variable LS  equation from the full three-dimensional LS equation.
           Sect. III defines and elaborates  the Bateman method for the two-variable LS equation.
            In Sect. IV,  the treatment of the singular integrals,
             the selection  of the interpolation  and quadrature grids, and the connection
             between Nystrom and Bateman approaches is presented.
               In Sec. V,    the
               results of calculations for a model two-nucleon potential are
               discussed and compared to benchmark results of Nystrom calculations.
               In Sec.VI, we summarize our conclusions.

    \begin {center} {\bf II. LIPPMANN-SCHWINGER EQUATION } \end{center}

\noindent The Lippmann-Schwinger (LS) equation for two-body
scattering in operator form reads
\begin{equation}
T(z)\ =\ V\ + V\, G_0(z)\, T(z)\ ,
\end{equation}
 where $T$ is the transition operator, $V$  the two-body  potential,
 $G_0\,=(z-H_0)^{-1}$, with  $H_0$ being the free hamiltonian and $z$
  the (complex) energy of the two-body system.
  Working in  the center-of-mass frame, the eigenstates of $H_0$
  are the relative momentum states $|\bf q>$,
  viz., $H_0|{\bf q}>\, =(q^2/2\mu)|{\bf q}>\, $.
  For on-shell scattering,  $z=E+i0$,  with $E=q_0^2/2\mu$,
  where $\mu$ is the reduced mass.
  The momentum-space matrix elements 
  $T({\bf q},{\bf q}_0\, ; z)\, (\equiv   <{\bf q}|T(z)|{\bf q}_0>\, $)
    satisfy  the three-dimensional integral equation
 \begin{equation}
 T ({\bf q},\, {\bf q}_0 \, ;z)\ =\ V\, ({\bf q}, \, {\bf q}_0 )\, + \,  \int \, {\mbox{d} }{\bf q}' \,
 \frac{ V \, ({\bf q},{\bf q }' ) \ T ({\bf q}', {\bf q}_0 \, ; z ) } { z\,  -\, q^{'2}/2\mu }
\end{equation}
 The $z$-dependence of the $T$-matrix
elements  $T ({\bf q},\, {\bf q}_0 \, ;z)$ will be suppressed,
unless  there is a  need to explicitly show the
energy dependence. The momentum-space representation $V\, ({\bf
q}, \, {\bf q}' )\,$
 of the potential $V$ is given as
 \begin {equation}
 V ({\bf q}, \, {\bf q}' )\,= \ <{\bf q}|V|{\bf q}'> \ = \,  \int \, d{\bf r}\, <{\bf q}|{\bf r}>\, V({\bf r})
 \, <{\bf r}|{\bf q}'> \, ,
 \end{equation}
 with $<{\bf r}|{\bf q}> \, =\,  e^{i {\bf r}\cdot {\bf q} }/(2\pi)^{3/2}$.
 For central potentials,
  $V ({\bf q}, \, {\bf q}' )$ and
 $T({\bf q}, \, {\bf q}' )$   depend
 only on $\, q,\, q'\, $ and  $x_{qq'}$. Here, $x_{qq'}$ denotes
 the cosine of the angle between vectors $\bf q$ and $\bf q'\, $.
 We denote the polar and azimuthal angles of the momentum vectors
 $\bf q$  by $\theta$ and $\phi$, respectively.
 We then
 have $\, x_{qq'}\, =\,{\hat {\bf q} }\cdot {\hat {\bf q}'}\,
 =\,  cos\,\theta_{qq'}\,=\, xx'+ ss'\, cos\, ( \phi-\phi')$,
  where $x=cos \, \theta $ and $s={\sqrt {1-x^2}}$.
 To emphasize this functional dependence on  $x_{qq'}$, we will occasionally use the notation
  $T(q,q', x_{qq'})$  to stand for  $T({\bf q}, {\bf q}'\,)\,$.

  For central potentials,
  the azimuthal-angle dependence in Eq. (2)
  can be integrated out
  to obtain a two-dimensional integral equation [1]. Towards this end,
  we introduce the  averaged momentum
  states $\, |qx>\, $ via
  \begin {equation}
  |qx>\,\,  =\,
  (2\pi)^{-1/2}\, \int_{0}^{2\pi}\, d\phi \,\, |{\bf q}>\, =
   \, (2\pi)^{-1/2}\, \int_{0}^{2\pi}\, d\phi \,\, |q\theta\phi>\, .
  \end{equation}
  For a two-body operator $A$, we introduce reduced matrix
  elements by
\begin{equation}
  A(q,x;q',x')\,  =\,  <qx|A|q'x'> \, =\,    \int_0^{2\pi} \, d\phi\,\,  A({\bf q},{\bf q}')\,
  =\, \int_0^{2\pi} \, d\phi\,\,  A(q,q',x_{qq'})\,  .  \\
 \end{equation}
For a rotationally invariant operator $A$,  the above integral is
independent of the variable $\phi'$. Integrating Eq. (2) over
$\phi$, we
  obtain  the two-variable  LS   equation for the reduced $T-$matrix:
  \begin{eqnarray}
  T(q,x;q_0,x_0) & = & V(q,x;q_0,x_0)\, \nonumber \\
                 &  &  + \,2\mu \int_{0}^{\infty}  q'^2dq'  \int_{-1}^{1}  dx'
   \frac {V(q,x;q',x')\, T(q',x';q_0,x_0) } {q_0^2-q'^2 +i0}\,  .
  \end{eqnarray}

  If we take the initial momentum vector ${\bf q}_0$
  along the $z$-axis,  the half-off-shell $T$-matrix element for
   a general final momentum vector $\bf q_f$ is then given by
  \begin {displaymath}
  <{\bf q_f}|T|q_0\hat {\bf z}>\,  =\, T (q_f,q_0,x_f)\,
   =\,  (2\pi)^{-1}\,  T(q_f,x_f\, ;q_0, 1)\, . \nonumber
  \end{displaymath}

  Direct  numerical solution of this two-variable
  Lippmann-Schwinger (LS) equation
   without invoking the partial wave expansion can be performed
   nowadays in commonly available computational platforms.
   The  approach used most frequently is the
   Nystrom method [14] in which
  the integrals over $q'$ and $x'$ are approximated
  by a suitable two-variable quadrature  and then   $x$ and $q$ variables are
  collocated at the quadrature points.
  This gives rise to a sytem of linear equations.
   Although the number of linear equations
   for the two-variable case is manageable and does not
    require special computing environment, going beyond
   two variables makes the matrix size  quickly become
   computationally prohibitive. Therefore, in the contexts of
   three and four body problems, alternatives to Nystrom method
   would be welcome. Bateman method described in the next
   section is such an alternative.\\

\begin {center} {\bf III. BATEMAN METHOD } \end{center}

Bateman method is based on a special kind of interpolation
of $V(q,x;q',x')$ on a finite set of grid points (nodes)
in the $q-x$ domain. Suppose two sets of nodes have been prescribed:
$N$ distinct points $\{q_1,q_2,...,q_N\}$
for $q$ in the interval $[0,\infty)$ and $M$ distinct points
$\{x_1,x_2,...,x_M\}$ for $x$ in the interval $[-1,+1]$.
The Cartesian grid generated by the Cartesian product
of these two sets is the set\\
\begin{eqnarray}
{\cal N}\  & = & \{q_1,q_2,...,q_N\}\times  \{ x_1,x_2,...,x_M \} \\
           & =  & \{ (q_n,x_m)\, : \,
                 1 \leq n \leq N,\  1 \leq m \leq M\}\, .
  \end{eqnarray}
  The set ${\cal N}$ will be referred to as the {\it interpolation grid}.
  The Bateman interpolate $V^B(q,x;q',x')$ of $V(q,x;q',x')$
  is defined as
  \begin{equation}
   V^B(q,x;q',x')\, =\, \sum_{n=1}^N \sum_{m=1}^M \sum_{n'=1}^N \sum_{m'=1}^M\
  V (q,x; q_n,x_m)\, {\bf \Lambda}_{nm,n'm'}\, V(q_{n'},x_{m'}; q',x')\, ,
  \end{equation}
  where the matrix ${\bf \Lambda}$ is defined via
  \begin{equation}
  ({{\bf \Lambda}^{-1}})_{nm,n'm'}\ =\ V(q_n,x_m;q_{n'}x_{m'})\, .
  \end{equation}

  The (exact) transition operator $T^B$
  for the separable potential $V^B$ is then
  given as
  \begin{equation}
    T^B(q,x;q',x')\, =\, \sum_{n=1}^N \sum_{m=1}^M \sum_{n'=1}^N \sum_{m'=1}^M\
  V (q,x; q_n,x_m)\, {\bf D}_{nm,n'm'}\, V(q_{n'},x_{m'}; q',x')\, ,
  \end{equation}
  where the matrix ${\bf D}$ is defined via
  \begin{eqnarray}
  ({{\bf D}^{-1}})_{nm,n'm'} & = & <q_nx_m|V-VG_0V|q_{n'}x_{m'}>\, , \nonumber \\
                             & = & V(q_n,x_m;q_{n'}x_{m'}) \nonumber \\
                             &   &   -  2\mu \int_{0}^{\infty}   q'^2dq' \int_{-1}^{1} dx'
   \frac {V(q_n,x_m;q',x')\, V(q',x';q_{n'},x_{m'}) } {q_0^2-q'^2 + i0}\ .
  \end{eqnarray}
This result corresponds to another instance of the weighted-residual approach [13]
 for the solution
 of the LS equation.  It also follows from Schwinger variational method [23,25,26]
 if the
wave function is expanded in the set of
reduced momentum states $\{|q_nx_m>\, , n=1,...,N\, , \, m=1,...,M\, \}. $
 On the other hand,  $V^B$ and $T^{B}$ can as well  be viewed, in the terminology of Ref. [27],
 as {\it inner-projection} approximations to the  operators $V$ and
 and $V(V-VG_0V)^{-1}V\, $[26], respectively.

\begin {center}
{\bf IV.  COMPUTATIONAL IMPLEMENTATION }
\end{center}
\begin {center}
{\bf A. Singular Integrals }
\end{center}
\noindent To computationally implement the Bateman method, we need to
introduce a quadrature rule on the $q-x$ computational domain  for
the evaluation of the matrix elements
$<q_nx_m|VG_0V|q_{n'}x_{m'}>\, $. We opt for a
tensor-product quadrature scheme.   Suppose $\{q_{\alpha}, \,
\alpha=1,...,N_q\}$ denote a suitable set of quadrature points for
the $q$-variable, with corresponding weights $\{w_{\alpha}, \,
\alpha=1,...,N_q\}$. Similarly, let $\{x_{\beta}, \,
\beta=1,...,N_x\}$ denote a  set of quadrature points for the
$x$-variable, with  corresponding weights $\{\rho_{\beta}, \,
\beta=1,...,N_x\}$. (Quadrature points are indexed by Greek letters $\alpha$ and $\beta$, while
indices $n$ and $m$ are reserved for interpolation points.)
 A quadrature rule of order
 $N_qN_x$ is thus provided by the   set   $ \{ (q_{\alpha},x_{\beta})\, \} $ of
 quadrature points (referred to as the {\it quadrature grid}) and the set
$ \{ (w_{\alpha},\rho_{\beta} )\, \}$  of quadrature weights.

  The evaluation of the  matrix elements
     $\, <q_nx_m|VG_0V|q_{n'}x_{m'}>$ are carried out using
     essentially the same subtraction procedure described in
     detail in Ref.[13]. The singular integral is separated into
     its real and imaginary parts as
     \begin{eqnarray}
      <q_nx_m|VG_0V|q{n'}x_{m'}>\,= \,2\mu \ {A}_{nm,n'm'} \,
       -\, i\pi \mu q_0 \, B_{nm,n'm'}(q_0) \, ,  \nonumber
      \end{eqnarray}
      where
      \begin{eqnarray}
       A_{nm,n'm'} \, & =   & \, {\cal P}\int_0^{q_{max}} \,  dq
       \, \frac {\,q^2\, B_{nm,n'm'}(q)} { q_0^2-q^2 \, } \nonumber \\
       B_{nm,n'm'}(q)\,&  =\, &
      \int_{-1}^{1}\, dx\,
       <q_nx_m|V|qx>\,   <qx|V|q_{n'}x_{m'}>\, ,\nonumber
      \end{eqnarray}
      where ${\cal P}$ stands for principle-value integral.
      By adding and  subtracting   a singular integral
      that can be evaluated analytically,
        singular term  $A_{nm,n'm'}$  is rearranged as  a sum of
        non-singular and singular terms:
      \begin{eqnarray}
              A_{nm,n'm'} & =  & A^{(ns)}_{nm,n'm'}\, +\, A^{(s)}_{nm,n'm'}\ , \nonumber
      \end{eqnarray}
      where
      \begin{eqnarray}
      A^{(ns)}_{nm,n'm'}  & =  & \int_0^{q_{max}}\,  dq\
      \frac { q^2 \,  B_{nm,n'm'}(q)\,   -\, q_0^2 \,  B_{nm,n'm'}(q_0) }
       { q_0^2-q^2  }\, ,  \nonumber \\
       A^{(s)}_{nm,n'm'}  & =  & B_{nm,n'm'}(q_0)\,
       \int_0^{q_{max}}\, dq\  \frac{q_0^2}{ q_0^2-q^2} \,
        =\, B_{nm,n'm'}(q_0) \,\frac{q_0}{2} \ln \frac{q_{max}+q_0}{q_{max}-q_0}\ . \nonumber
     \end{eqnarray}
      As the integrals involved  in $B_{nm,n'm'}(q)$ and $A^{(ns)}_{nm,n'm'} \,$
      are non-singular, they are amenable to
      approximation  by quadrature with the result
     \begin{eqnarray}
     B_{nm,n'm'}(q)   & \approx   & \Sigma_{\beta=1}^{N_x}\, \rho_{\beta}\,
      <q_nx_m|V|q x_{\beta}>\, <q x_{\beta}|V|q_{n'}x_{m'}>\,
      \nonumber \\
      A_{nm,n'm'} & \approx   & \Sigma_{\alpha=1}^{N_q}\, \, w_{\alpha}\,
       q_{\alpha}^2\,
      \frac {B_{nm,n'm'}(q_{\alpha} )}
      { q_0^2-q_{\alpha}^2 \, } \, +\, C_{sing}\, q_0^2\, B_{nm,n'm'}(q_0) \nonumber
      \end{eqnarray}
      \noindent  where
      \begin{displaymath}
      C_{sing}\,= \,\,\frac{1}{2q_0} \ln {\frac{q_{max}+q_0}{q_{max}-q_0}} \, -
       \, \sum_{\alpha=1}^{N_q}\, \frac{w_{\alpha}}{q_0^2-q_{\alpha}^2}\, .
      \end{displaymath}
   Note that as long as the  quadrature grid treats the singularity symmetrically and
   is sufficiently dense in its vicinity, the correction term involving $C_{sing}$
   comes out to be very small and can even be omitted without causing a significant loss of
   accuracy.

    \begin {center}
{\bf B. Interpolation Grid and Quadrature  }
\end{center}

    \noindent To construct the Cartesian interpolation grid, we need to specify
     two uni-variate grids,  one in $q$ and one in $x$.
     To define the $q$-grid $\{ q_n \}\, $,
     the full $q$-domain is divided into two intervals:
     $[0,\, 2q_0]$,  and $[2q_0, \infty )$.
     This scheme is adopted to
     treat the singularity of the LS kernel at $q'=q_0$ in Eqs. (2) or(6)
      as symmetrically as possible, and to have a $q$-grid that is denser
      in the vicinity of $q_0$.
     To this end, the first interval $[0,\, 2q_0]\, $,
     is   subdivided into $I_1$  (equal) subintervals (finite
     elements).

     The  second interval
     $[2q_0,\infty)$, however,  is first  mapped to $[-1,+1]$
     via the  transformation
     \begin{equation}
     u\, =\,  \frac{q-2q_0-f}{q-2q_0+f} \, , \ \  \mbox{or}\ \  q\, =\, 2q_0\,  +\, f\, \frac {1+u}{1-u}\ ,
     \end{equation}
      where $f$ is a scale factor.
       The $q$-variable is cut off
      at some large but finite value $q_{max}$ by adopting an upper limit $u_{max}\ (<1)$ to the variable $u$.
      In the calculations reported in this paper, we used $u_{max}=0.99$,
      which corresponds   to $q_{max}$ values of
       several thousand. This variable transformation
        is instrumental  for a discretization of the
         semi-infinite interval
      $[0,\infty)$ with relatively  few finite elements.
       The interval $[-1,u_{max}]$ is divided
       into $I_2$ equal finite elements. Note that this gives rise
        to  a non-uniform partitioning for the $q$-variable in the interval $[2q_0, q_{max}]$.
        The total number of finite elements
       covering the computational interval
    $[0,q_{max}]$ is $I \ (\equiv I_1+I_2) $.
     The choice $I_2=3I_1$ (hence $I=4I_1)$ was found adequate after
     some experimentation. The $q$-grid
     $\{ q_1, q_2,...,q_N \}$ consists of the  end points
     and mid-points of this
     finite-element partitioning.
     Note that $N=2I+1$,  $q_1=0$,  and $q_N=q_{max}$.

       The specification of the $x$-grid  proceeds similarly
       to that of the $q$-grid.
       The interval $[-1,1]$ is
       partitioned into $J$ equal subintervals (finite-elements).
    Collecting and ordering
   the endpoints and midpoints of the finite elements together, we  define
    the set of grid points
   $\, \{ {x_{1},x_{2},..., x_{M}}\,  \}\, ,$ where
   $\, M\, =\, 2J+1\, $,  $\, x_{0}\, =\, -1 \, $ and $\, x_M\, =\, +1 \, $.

    To construct a composite Gauss-Legendre quadrature rule for $q$,
      each finite-element  $[q_{2i-1},q_{2i+1}]\, , i=1,2,...,I_1\,$, in the interval $[0,2q_0]$,
      is mapped to $[-1,1]$
       via $s=(2q-q_{2i-1}-q_{2i+1})/(q_{2i+1}-q_{2i-1})\,$. For finite
       elements in $[2q_0,q_{max}]$, we map the  finite elements
       $[u_{2i-1},u_{2i+1}]\,$, for $i=1,...,I_2\, $,
       into $[-1,+1]$ via the map $s=(2u-u_{2i-1}-u_{2i+1})/(u_{2i+1}-u_{2i-1})\,$.
    We choose a set of  $n_q$ Gauss-Legendre quadrature points in the local variable $s$,
    and then transform them  back to $q$-variable.
   The Gauss-Legendre quadrature points   for all elements
   are then combined and ordered to form a composite quadrature
   rule with
   the set of quadrature points
   $\, \{ \, q_{\alpha}  , \,  \alpha\, = \, 1,2,...,N_q\, \} \, $,
   where
    $N_q\, = I\,n_q $.
    The quadrature weights are similarly collected in the set
    $\, \{ w_{\alpha}  , \,  \alpha\, =\, 1,2,...,N_q\, \}\, $.
    In the calculations reported in the next section,
    values of $n_q$  ranged  from 4 to 32, depending on the
    fineness of the finite-element partitioning. To ensure results stable
    within 6-7 digits after the decimal point, the total number $N_q$
     of quadrature points were typically in the order of 160-200,
     although 3-4 digit accuracy could be achieved with, say, $N_q=64$.

    For the $x$-variable,  each finite element
     $\, [\, x_{2j-1},x_{2j+1}\, ] \, $,
     is mapped to $[-1,+1]$ via the map
      $s=(2x-x_{2i-1}-x_{2i+1})/(x_{2i+1}-x_{2i-1})\, $ for
      $i=I_1,..., I\, $.
     We choose $n_x $
    Gauss-Legendre quadrature points in $s$ and transform back to the $x$-variable.
    The quadrature points and their weights over individual finite elements
    are collected
     in the sets
    $\{ x_{\beta}\, , \,  \beta\, =\, 1,2,...,N_x\, \}  $, and
    $\{ \rho_{\beta}\, , \, \beta \, =\, 1,2,...,N_x\, \} $,
    where $N_x=J\,n_x\,$. In our calculations, typically $N_x=80$
    was sufficient to obtain results stable within 6 digits.\\

       \begin {center}
  {\bf C. Connection between Nystrom and Bateman Methods }
  \end{center}

       The reference results against which the results of
      Bateman method will be tested
   are  obtained by solving the two-variable integral equation
   via the Nystrom method. Details of the computational implementation 
   of the Nystrom method is given in [13].  
   Nystrom result  for  T-matrix elements on the quadrature grid 
    can be recast as 
   
    \begin {eqnarray}
      T^{N}(q_{\alpha},x_{\beta};q_{{\alpha}'},x_{{\beta}'})\, =\,
       \hspace*{8.2cm} \nonumber \\
        \ \  \sum_{{\alpha}''=1}^{N_q}  \sum_{{\beta}''=1}^{M_x}
    \sum_{{\alpha}'''=1}^{N_q} \sum_{{\beta}'''=1}^{M_x}\
  V (q_{\alpha},x_{\beta}; q_{{\alpha}''},x_{{\beta}''})\, {\bf D}_{{\alpha'' \beta''},{\alpha}'''{\beta}'''}\,
  V(q_{{\alpha}'''},x_{{\beta}'''}; q_{{\alpha}'},x_{{\beta}'})\, ,
  \end {eqnarray}
  \noindent where the matrix ${\bf D}$ is given as
  \begin{equation}
  ({{\bf D}^{-1}})_{ {\alpha}{\beta},{\alpha}'{\beta}' }  =
                    <q_{\alpha},x_{\beta}|V-VG_0V|q_{{\alpha}'},x_{{\beta}'}>\, .
  \end{equation}
 In Eq. (15),  the multivariable integrals implicit
 in $ <q_{\alpha},x_{\beta}|VG_0V|q_{{\alpha}'},x_{{\beta}'}>\, $
 are,  of course, evaluated using the quadrature grid
 $(q_{\alpha},x_{\beta})$ with weights
 $(w_{\alpha},\rho_{\beta})$. 

 Eq. (14) represents a somewhat unusual depiction of
 the   Nystrom method.  
  This form manifestly shows  the connection between Nystrom and Bateman
   methods.   Eq. (14) is in fact the Bateman expression
   when quadrature grid also doubles
   as the interpolation grid.  Therefore,    if a given grid
   is  employed  as both interpolation and quadrature grids in the Bateman method,
      then the ensuing Bateman calculation  becomes  equivalent
      to the Nystrom calculation (using the same quadrature rule).
      Stated differently, a Nystrom calculation can indeed
      be viewed as  a Bateman calculation as well. This formal equivalence
       is borne out in computations.
      Although computational organizations of
      our Nystrom and Bateman codes  are
      quite different, their results agree to at least 8 digits
      when interpolation and quadrature grids are chosen to coincide,
      giving confidence that computer codes
      are performing satisfactorily.

      \begin {center} {\bf V. RESULTS  } \end{center}

 We have tested the Bateman method for the Malfliet-Tjon III
 ( MT-III)  model [24] for the two-nucleon potential:
      \begin{displaymath}
        V(r)\, =\, V_R\, e^{-\mu_Rr}\, - \, V_A\, e^{-\mu_Ar}\,
     \end{displaymath}
     whose momentum-space  representation is given as
     \begin{displaymath}
        V({\bf q},{\bf q}')\,   =\, \frac{1}{2\pi^{2}}\,\left(  \frac{V_R}{({\bf q}-{\bf q}')^2+\mu_R^2}
        \,  - \, \frac{V_A}{({\bf q}-{\bf q}')^2 +\mu_A^2 } \, . \right)
     \end{displaymath}

     For this potential the azimuthal integration in Eq. (4) can be carried out analytically to give

\begin{eqnarray}
        V(q,x;q',x')\,  & = & \,    \frac{V_R/ \pi} { \sqrt {(q^2+q'^2-2qq'xx'+\mu_R^2)^2
         - 4 q^2q'^2(1-x^2)(1-x'^2) } } \nonumber \\
        \,& \ & - \,  \frac{V_A/\pi} { \sqrt {(q^2+q'^2-2qq'xx'+\mu_A^2)^2
         - 4 q^2q'^2(1-x^2)(1-x'^2) } }   \nonumber
  \end{eqnarray}
  The parameters for MT-III potential are taken from  Ref. [5]: $V_A=626.8932$ MeV fm,
  $V_R=1438.723$ MeV fm, $\mu_A=1.55$ fm$^{-1}$ and $\mu_R=3.11$ fm$^{-1}$.
  For the two-nucleon calculations,
  we set nucleon  mass
   and $\hbar$ to unity and take $fm$ as the unit of length.
   The nucleon mass adopted yields the conversion factor
    $1 fm^{-2}=41.47$ MeV.

  For general potentials,  $V(q,x;q',x')$ may  not be available analytically. Its numerical
  generation by applying a suitable quadrature to the integral over the azimuthal angle $\phi$
   is quite feasible. In fact, we have  tested
   this aspect on the present potential.  Using a  composite 64-point Gauss-Legendre rule for
   the $\phi$-integral,  the results  were indistinguishable within 7-8 digits from those of
   the analytical reduced potential.

   Tables I-IV  show the convergence pattern of   the Bateman results  as
    the number of grid points in $q$ and $x$ variables are increased.
    Two collision energies considered are
    $E=150 \, MeV$ and $E=\, 400\, MeV$. Shown are the on-shell T-matrix elements
   $T(q_0,q_0,x)\,( \equiv T(q_0,x;q_0,1)/2\pi \, )$,
   in units of $MeV-fm^3\,$,  for three values of $x$.
   Also shown is the s-wave component of
   the on-shell T-matrix, obtained by numerically averaging $T(q_0,q_0,x)$ over $x$.
    Reference values in these tables were
  obtained using $N_q=200$ and $N_x=80$ in the Nystrom method. These are stable
  to within at least the number of digits shown
  against further increases in  the computational parameters
  like $N_q$, $N_x$, $q_{\max} $ and
  against the variations in the  the distribution pattern of quadrature
  points  in the $q-x$ plane.

  An examination of these tables show that results
  accurate to 3-4 digits can be obtained with relatively
  few grid points. However, going beyond this level of  accuracy may
  require much finer interpolation grids. Especially, the peak at
  the forward direction requires  many grid points to reach 6
  digit accuracy. For directions other than forward, convergence is
  rather rapid. As a remedy for the slow convergence around the forward direction,
  a non-uniform $x$-grid with more points around
  $x=1$ might be considered.

  Whenever the number of interpolation
  points are comparable to the number of quadrature points,
  the  Bateman and Nystrom
  methods will yield similar levels of accuracy. In fact,
  as mentioned earlier,
  when a quadrature grid is also used as interpolation grid,  Bateman method
  degenerates into the Nystrom method. Conversely, a Nystrom calculation is
  at the same time a Bateman
  calculation. In a sense, Nystrom is a restricted type of Bateman
  method in which interpolation and quadrature grids are the same.
  That Bateman approach distinguishes between interpolation and
  quadrature grids is a strength of the Bateman approach over the
  Nystrom approach. For a given (crude) interpolation grid,
   the integral involved in the
   matrix element $\,\,  <q_nx_m|VG_0V|q_{n'}x_{m'}>$ can be
  calculated with a finer quadrature grid, without affecting the
  the order of the matrix ${\bf D}$. In contrast, in Nystrom
  method, the quadrature grid used to discretize the integral will
  have to  be used as collocation points in order to obtain a
  consistent set of equations.

    \begin {center} {\bf VI. DISCUSSION and CONCLUSIONS  } \end{center}

    We have shown that the  multi-variate Bateman interpolation
    of $V(q,x;q',x')$ on a grid provide a simple and viable
    computational scheme to solve the LS equation
    without invoking angular momentum decomposition.
    In terms of computational
     complexity, Bateman approach  is quite
     similar to the Nystrom method. For a given quadrature grid, both methods
     involve essentially the same sampling of the potential and the free propagator,
     but
     In fact, when interpolation
     and quadrature grids are taken to coincide, Bateman and
     Nystrom methods become equivalent.

     The use of two grids in
     the Bateman approach, one for interpolation/collocation and one for
     quadrature evaluation of matrix elements, gives it an
     additional flexibility. The matrix dimension is determined by
     the interpolation grid. However, calculation of the integral
      in Eq. (10) for the construction of the matrix ${\bf D}^{-1}$  can be
     carried out  with a higher-order quadrature rule
      without affecting the matrix size. From this point of view,
      Nystrom approach is a restricted type of Bateman
      method with the same grid used
      for the purposes of interpolation and quadrature both.
      Thus, Bateman  can be expected to be more effective
       for small interpolation grids  than the Nystrom
      method (with a quadrature grid of similar size).

      We find that Bateman method can yield 3-4 digit accuracy
      with relatively small numbers of interpolation points.
      However, as the size of the interpolation grid is increased to achieve
       higher level of accuracy, the advantage associated with the reduction
       in  matrix dimension may disappear to some extent. With large sets of
       interpolation and quadrature points,
       both Nystrom and Bateman methods  are
       expected to perform  satisfactorily.

         In three- and four-particle contexts, the two-particle
     $T$-matrix $<{\bf q}|T(E)|{\bf q}'>$  is needed
     for many different two-particle energies $E$ and
      off-shell momenta ${\bf q}$
     and ${\bf q}'$. Bateman method could provide an effective
     way of generating arbitrary off-shell $T$-matrix elements
     needed in direct momentum-vector approaches
      to solve  three-particle Faddeev equations
      without employing partial-wave decomposition [28,29].

\begin {center} {\bf REFERENCES  } \end{center}

\begin{enumerate}

    \item Ch. Elster, J. H. Thomas, and W. Glockle,
  Few-Body Syst. {\bf 24}, 55 (1998).

\item W. Schadow, Ch. Elster,  and W. Glockle,
  Few-Body Syst.  {\bf 28}, 15 (2000).

\item J. Shertzer and A. Temkin, Phys. Rev. A {\bf 63}. 062714
(2001).

\item G. L. Caia, V. Pascalutsa and  L. E. Wright, Phys. Rev. C
{\bf 69}, 034003 (2004).

\item B. M. Kessler, G. L. Payne, and W. N. Polyzou, Phys. Rev. C
{\bf 70},  034003 (2004).

\item A. S. Kadryov, I. Bray, A. T. Stelbovics, and B. Saha,
 J. Phys. B {\bf 38},  509 (2005).

\item G. Ramalho, A. Arriaga, and M. T. Pena,
 Few-Body Syst.  {\bf 39}, 123 (2006).

\item  M. Rodriguez-Gallardo, A. Deltuva,  E. Cravo,
  R. Crespo, and A. C. Fonseca,
Phys. Rev. C  {\bf 78},  034602 (2008).

\item  M. Rodriguez-Gallardo, A. Deltuva,
  R. Crespo, E. Cravo, and A. C. Fonseca,
Eur. Phys. J A  {\bf 42},  601 (2009).

\item A. S. Kadyrov, I. B. Abdurakhmanov, I. Bray,  and A. T.
Stelbovics, Phys. Rev. A  {\bf 80}, 022704 (2009).

\item   J. Golak,  R. Skibinski,  H. Witala,  K. Topolnicki,
 W. Glockle,  A. Nogga, and  H. Kamada,
  Few-Body Syst.  {\bf 53},  237(2012).

 \item S. Veerasamy, Ch. Elster,  and W. N. Polyzou,
 Few-Body Syst. (2012). doi:10.1007/s00601-012-0476-1

 \item Z. C. Kuruoglu,
  Few-Body Syst. (2013). doi:10.1007/s00601-013-0732-z.

   \item K. E. Atkinson, {\it  A Survey of Numerical Methods for the Solution of
  Fredholm Integral Equations of the Second Kind}. (SIAM, Philadelphia, 1976)

 \item H. Bateman, Proc. R. Soc. Lond. A {\bf 100}, 441 (1922).

 \item G. T. Thompson, J.Assoc. Comput. Mach, {\bf 4}, 314 (1957).

 \item S. Joe and I. H. Sloan, Numer. Math. {\bf 49}, 499(1986).

 \item V. B. Belyaev and A. L. Zubarev,
   Sov. J. Nucl. Phys. {\bf 14}, 305 (1972).

   \item V. I. Kukulin,  Sov. J. Nucl. Phys. {\bf 14}, 481 (1972).

    \item S. Oryu, Prog. Theor. Phys. {\bf 52}, 550 (1974)

 \item V. B. Belyaev, E. Wrzecionko, and B. F. Irgaziev,
  Sov. J. Nucl. Phys. {\bf 20}, 664(1975)

  \item T. K. Lim and J. Giannini, Phys. Rev. A {\bf 18}, 517 (1978).

  \item A. L. Zubarev,
   Sov. J. Part. Nucl. {\bf 9}, 188 (1978).

   \item R. A. Malfliet and J. A. Tjon, Nucl.Phys. A {\bf 27},161 (1969).

  \item S. K. Adhikari, {\it Variational Principles and the Numerical Solution
   of Scattering Problems} ( Wiley, New York, 1998).

  \item Z. C. Kuruoglu and D. A. Micha, J. Chem. Phys. {\bf 72}, 3328 (1980).

   \item  P. O. L\"owdin, {\it Linear Algebra for Quantum Theory}.
    (Wiley, New York, 1998)

 \item H. Liu, Ch. Elster, and  W. Glockle,
 Phys. Rev. C  {\bf 72},  054003 (2005),

 \item  Ch. Elster,  W. Glockle, and H. Witala,
   Few-Body Syst.  {\bf 45}, 1 (2009).

\end{enumerate}

\newpage

\noindent TABLE\ I. Convergence of the Bateman method with respect
to the number points in the $q$-grid. Shown are the
on-shell T-matrix elements $T(q_0,q_0,x)$  for $q_0=1.901860 \, fm^{-1}$
 or $E=150$ MeV. Parameters $N$ and $M$ denote the number of
points in the $q$- and $x$-grids, respectively.\\

\begin{tabular}{ccllll}
\hline\hline

\multicolumn{1}{c}{$\ N\ $ } & \multicolumn{1}{c}{$\ M\ $ } &
 \multicolumn{1}{l}{s-wave}
 &\multicolumn{1}{l}{ x=+1.0 } &
\multicolumn{1}{l} { x=0.0 }
&\multicolumn{1}{l}{ x=-1.0 } \\
\hline \hline
\multicolumn{2}{c}{ } & \multicolumn{4}{c} {Real part of $T(q_0,q_0,x)$}\\
\cline{3-6}
 9   & 41 & 0.102424 & -6.012713 & 0.478064 & 0.232340\\
 17  & 41 & 0.103792 & -6.084988 & 0.490515 & 0.234057\\
 25  & 41 & 0.103967& -6.092244 & 0.491699 & 0.233998 \\
 33  & 41 & 0.103982 & -6.092730 & 0.491771 & 0.233959 \\
 41  & 41 & 0.103982 & -6.092758 & 0.491778 & 0.233969 \\
      & 51 & 0.103986 & -6.092771 & 0.491771 & 0.233962 \\

 49  & 41 & 0.103984 & -6.092766 & 0.491774 & 0.233962 \\

 65  & 41 & 0.103984 & -6.092765 & 0.491774 & 0.233962 \\

 81  & 41 & 0.103983 & -6.092763 & 0.491776 & 0.233967 \\
      & 51 & 0.103986 & -6.092774 & 0.491771 & 0.233959 \\
\hline
 \multicolumn{2}{l}{Nystrom} & \multicolumn{1}{l}{0.103988 } &
\multicolumn{1}{l}{-6.092782} & \multicolumn{1}{l}{0.491768}
& \multicolumn{1}{l}{0.233958 }\\
\hline \hline
\multicolumn{2}{c}{ } & \multicolumn{4}{c} {Imaginary part of $T(q_0,q_0,x)$}\\
\cline{3-6}
 9   & 41 & 0.0207717 & -1.837919 & 0.272400 & 0.329956\\
 17  & 41 & 0.0213556 & -01.927475 & 0.284770 & 0.362303\\
 25  & 41 &  0.0214309 & -1.936595 & 0.286015 & 0.365442\\
 33  & 41 &  0.0214373 & -1.937290 & 0.286094 & 0.365635\\
 41  & 41 & 0.0214371 & -1.937226 & 0.286101 & 0.365647 \\
     & 51 & 0.0214389 & -1.937237 & 0.286098 & 0.365647 \\
 49  & 41 & 0.0214383 & -1.937230 & 0.286099 & 0.365643 \\
 \
 65  & 41 & 0.0214381 & -1.937230 & 0.286100 & 0.365643 \\
 81  & 41 & 0.0214376 & -1.937231 & 0.286100 & 0.365647 \\
     & 51 & 0.0214391 & -1.937238 & 0.286098 & 0.365646 \\
\hline
 \multicolumn{2}{l}{Nystrom}
  & \multicolumn{1}{l}{0.0214399 }
   & \multicolumn{1}{l}{-1.937247}
   & \multicolumn{1}{l}{0.286097}
   & \multicolumn{1}{l}{0.365649}\\
\hline \hline
\end{tabular}

\newpage

\noindent TABLE II. Convergence of the Bateman method with respect
 to number of  points  in the $x$-grid. Shown are the
 on-shell T-matrix elements $T(q_0,q_0,x)$  for $q_0=1.901860 \,  fm^{-1}$
 or $E=150$ MeV. Parameters $N$ and $M$ denote the number of
 points in the $q$- and $x$-grids, respectively.\\

\begin{tabular}{ccllll}
\hline\hline
\multicolumn{1}{c}{$\ N\ $ } & \multicolumn{1}{c}{$\ M\ $ } &
 \multicolumn{1}{l}{s-wave}
 &\multicolumn{1}{l}{ x=+1.0 } &
\multicolumn{1}{l} { x=0.0 }
&\multicolumn{1}{l}{ x=-1.0 }
\\ \hline \hline
\multicolumn{2}{c}{ } & \multicolumn{4}{c} {Real part of $T(q_0,q_0,x)$}\\
\cline{3-6}
 41 & 21 & 0.103941 & -6.092561 & 0.491835 & 0.234068 \\

      & 31 & 0.103979 & -6.092735 & 0.491782 & 0.233966 \\

     & 41 & 0.103982 & -6.092758 & 0.491778 & 0.233969 \\

      & 51 & 0.103986 & -6.092771 & 0.491771 & 0.233962 \\

\hline
 \multicolumn{2}{l}{Nystrom} & \multicolumn{1}{l}{0.103988 } &
\multicolumn{1}{l}{-6.092782} & \multicolumn{1}{l}{0.491768}
& \multicolumn{1}{l}{0.233958 }\\
\hline \hline
\multicolumn{2}{c}{ } & \multicolumn{4}{c} {Imaginary part of $T(q_0,q_0,x)$}\\
\cline{3-6}
 41  & 21 & 0.0214173 & -1.937046 & 0.286109 & 0.365694\\

      & 31 & 0.0214358 & -1.937200 & 0.286102 & 0.365633 \\

      & 41 & 0.0214371 & -1.937226 & 0.286101 & 0.365647 \\

      & 51 & 0.0214389 & -1.937237 & 0.286098 & 0.365647 \\

\hline
 \multicolumn{2}{l}{Nystrom}
  & \multicolumn{1}{l}{0.0214399 }
   & \multicolumn{1}{l}{-1.937247}
   & \multicolumn{1}{l}{0.286097}
   & \multicolumn{1}{l}{0.365649}\\
\hline \hline
\end{tabular}

\newpage

\noindent TABLE III. Convergence of the Bateman method with
respect to the number of  points  in the $q$-grid. Shown are the
 on-shell T-matrix elements $T(q_0,q_0,x)$  for $q_0=3.105725\,  fm^{-1}$
 or $E=400$ MeV. Parameters $N$ and $M$ denote the number of
 points in the $q$- and $x$-grids,
 respectively.\\

\begin{tabular}{ccllll}
\hline\hline
\multicolumn{1}{c}{$\ N\ $ } & \multicolumn{1}{c}{$\ M\ $ } &
 \multicolumn{1}{l}{s-wave}
 &\multicolumn{1}{l}{ x=+1.0 } &
\multicolumn{1}{l} { x=0.0 }
&\multicolumn{1}{l}{ x=-1.0 }
\\ \hline \hline
\multicolumn{2}{c}{ } & \multicolumn{4}{c} {Real part of $T(q_0,q_0,x)$}\\
\cline{3-6}
  9    & 41 & -0.0927120 & -6.051275 & 0.472848 & 0.283085\\
 17  & 41 & -0.0800118 & -6.122230 & 0.453371 & 0.253529\\
 25  & 41 & -0.0783654 & -6.154789 & 0.454285& 0.249604 \\
 33  & 41 & -0.0781718 & -6.162070 & 0.454812 & 0.249213 \\
 41  & 41 & -0.0781400 & -6.163427 & 0.454910 & 0.249137 \\
     & 51 & -0.0781380 & -6.163477 & 0.454910 & 0.249148 \\
 49  & 41 & -0.0781348 & -6.163683 & 0.454928& 0.249129 \\
 65  & 41 & -0.0781353 & -6.163743 & 0.454934 & 0.249141 \\
 81  & 41 & -0.0781350 & -6.163750& 0.454934 & 0.249145 \\
   & 51 & -0.0781372 & -6.163762 & 0.454939 & 0.249155 \\
\hline

 \multicolumn{2}{l}{Nystrom}
 & \multicolumn{1}{l}{-0.0781299}
  &\multicolumn{1}{l}{-6.163808}
 & \multicolumn{1}{l}{0.454930}
& \multicolumn{1}{l}{0.249139 }\\
\hline \hline
\multicolumn{2}{c}{ } & \multicolumn{4}{c} {Imaginary part of $T(q_0,q_0,x)$}\\
\cline{3-6}
9  & 41 & 0.0293396 & -1.276232 & 0.079004 & -0.0885716\\
 17  & 41 & 0.0212913 & -1.288721 & 0.102911 & -0.0768172\\
 25  & 41 &  0.0203595 & -1.306253 & 0.109280 & -0.0773454\\
 33  & 41 &  0.0202510 & -1.310558 & 0.110482 & -0.0775932\\
 41  & 41 & 0.0202334 & -1.311349 & 0.110694 & -0.0776496\\
     & 51 & 0.0202332 & -1.311435 & 0.110704 & -0.0776371\\

 49  & 41 & 0.0202306 & -1.311405 & 0.110731& -0.0776569\\

 65  & 41 & 0.0202305 & -1.311531 & 0.110737& -0.0776534\\

 81  & 41 & 0.0202303 & -1.311535 & 0.110738& -0.0776498\\
     & 51 & 0.0202312 & -1.311613 & 0.110745& -0.0776478\\
\hline
 \multicolumn{2}{l}{Nystrom}
 & \multicolumn{1}{l}{0.0202292 }
 & \multicolumn{1}{l}{-1.311641}
 & \multicolumn{1}{l}{0.110753}
 & \multicolumn{1}{l}{-0.0776420}\\
\hline \hline

\end{tabular}

\newpage

\noindent TABLE IV. Convergence of the Bateman method with respect
 to the number of  points  in the $x$-grid. Shown are the
 on-shell T-matrix elements $T(q_0,q_0,x)$  for $q_0=3.105725\,  fm^{-1}$
 or $E=400$ MeV. Parameters $N$ and $M$ denote the number of
 points in the $q$- and $x$-grids, respectively.\\

\begin{tabular}{ccllll}
\hline\hline
\multicolumn{1}{c}{$\ N\ $ } & \multicolumn{1}{c}{$\ M\ $ } &
 \multicolumn{1}{l}{s-wave}
 &\multicolumn{1}{l}{ x=+1.0 } &
\multicolumn{1}{l} { x=0.0 }
&\multicolumn{1}{l}{ x=-1.0 }
\\ \hline \hline
\multicolumn{2}{c}{ } & \multicolumn{4}{c} {Real part of $T(q_0,q_0,x)$}\\
\cline{3-6}
41  & 21 & -0.0781765 & -6.164592 & 0.454996  & 0.249352 \\

    & 31 & -0.0781459 & -6.163433 & 0.454913 & 0.249137 \\

    & 41 & -0.0781400 & -6.163427 & 0.454910 & 0.249137 \\

     & 51 & -0.0781380 & -6.163477 &  0.454910 & 0.249148 \\

\hline

 \multicolumn{2}{l}{Nystrom}
 & \multicolumn{1}{l}{-0.0781299}
  &\multicolumn{1}{l}{-6.163808}
 & \multicolumn{1}{l}{0.454930}
& \multicolumn{1}{l}{0.249139 }\\
\hline \hline
\multicolumn{2}{c}{ } & \multicolumn{4}{c} {Imaginary part of $T(q_0,q_0,x)$}\\
\cline{3-6}
  41 & 21 &  0.0201645  & -1.309207  & 0.110317  & -0.0778120\\

     & 31 & 0.0202295 & -1.311040  & 0.110645  & -0.0776740 \\

     & 41 & 0.0202334 & -1.311349 & 0.110694  & -0.0776496\\

     & 51 & 0.0202332 & -1.311435 & 0.110704  & -0.0776371\\

\hline
 \multicolumn{2}{l}{Nystrom}
 & \multicolumn{1}{l}{0.0202292 }
 & \multicolumn{1}{l}{-1.311641}
 & \multicolumn{1}{l}{0.110753}
 & \multicolumn{1}{l}{-0.0776420}\\

\hline \hline
\end{tabular}

\end {document}